\documentclass[3p, times, review ]{elsarticle}

\usepackage[utf8]{inputenc}
\usepackage[english]{babel}
  \usepackage[]{algorithm2e}
\usepackage[utf8]{inputenc}
\usepackage[T1]{fontenc}
\usepackage{array}
\usepackage{amssymb}
\usepackage{amsfonts}
\usepackage{amsmath}
\usepackage{cases}
\usepackage{hyperref}
 
 \newcommand{\new}[1]{\textcolor{black}{#1}}
\usepackage{amsthm}
\usepackage{graphicx}
\usepackage{subfig} 
\usepackage{textcomp}
\usepackage{bm}
\usepackage{booktabs}
\usepackage{mathtools}

\makeatletter

\newcommand{\Rmnum}[1]{\expandafter\@slowromancap\romannumeral #1@}
\makeatother

\theoremstyle{definition}

\newcommand{\K}{\mathcal{K}}

\newcommand{\unit}[2]{#1\,\mathrm{#2}}

\usepackage{color}

\usepackage[draft,disable]{todonotes}

\newcommand{\RN}[1]{%
  \textup{\uppercase\expandafter{\romannumeral#1}}%
}
 
\usepackage[figuresright]{rotating}

\begin{document}

\begin{frontmatter}


\title{Frequency dependence of dielectrophoresis fabrication of single-walled carbon nanotube field-effect transistor}
\author[Shezan]{Yousef Adeli Sadabad}
 
\author[tuwien,Hannover]{Amirreza Khodadadian\corref{cor1}}
\ead{amirreza.khodadadian@tuwien.ac.at}
\author[Shezan]{Kiarash Hosseini Istadeh}
\author[Shezan]{Marjan Hedayati} 
 \author[Shezan]{Reza Kalantarinejad} 
\author[tuwien,ASU]{Clemens Heitzinger}
\cortext[cor1]{Corresponding author}

\address[tuwien]{Institute for Analysis and Scientific Computing, 
  Vienna University of Technology (TU Wien),
  Wiedner Hauptstraße 8--10, 
  1040 Vienna, Austria}
  \address[Hannover]{
  	Institute of Applied Mathematics, Leibniz University Hannover, Welfengarten 1, 30167
  	Hanover, Germany
  }
 
  \address[ASU]{School of Mathematical and Statistical Sciences, Arizona State University, Tempe, AZ 85287, USA}
  \address[Shezan]{Shezan Research and Innovation Center, No. 25, Innovation 2 St., Pardis TechPark, Tehran, Iran}

\date{}

\begin{abstract}
 In this paper, we present a new theoretical model for the dielectrophoresis (DEP) process of the single-walled Carbon nanotubes (SWCNT).  We obtain a different frequency interval for the alignment of wide energy gap semiconductor SWCNTs which shows a considerable difference with the prevalent model. For this, we study two specific models namely, the spherical model and the ellipsoid model to estimate the frequency interval. Then, we perform the DEP process and use the obtained frequencies (of spherical and ellipsoid models) for the alignment of the SWCNTs.
 Our empirical results declare the theoretical prediction, i.e., a crucial step toward the realization of carbon nanotube field-effect transistor (CNT-FET) with the DEP process based on the ellipsoid model.

\end{abstract}

\begin{keyword}
 dielectrophoresis, single-wall carbon nanotube, spherical model, ellipsoid model
\end{keyword}

\end{frontmatter}

\section{Introduction}\label{s:intro}
A graphene sheet rolled up into a cylinder named SWCNT is either metallic or semiconducting depending on its geometrical structure. An SWCNT is a material with unique properties, like high tensile strength, high electrical/thermal conductivity, high elastic modulus, and ductility. All these properties make CNT-based products to have high scientific importance in addition to a high potential of technical applications such as thermal conduction enhancement, composites, filtration, sensors, and microelectronics.

 Carbon nanotubes with their unique electronic and mechanical properties have drawn great attention in a broad range of applications especially in Nanoelectronics and sensing applications \cite{kalantari2010computational,zhang2008recent,sinha2006carbon}.
 Electronic devices fabricated using individual SWNT have shown outstanding device performance surpassing those of silicon. A critical step to obtain these practical devices is to deposit well organized and highly aligned CNTs in desired locations. There are a different number of methods to align CNTs such as chemical and biological patterning, Langmuir-Blodgett assembly, etc. The interested reader can refer to \cite{auvray2005chemical,keren2003dna,yu2007large,li2007langmuir,jin2004scalable,engel2008thin}. These methods have their shortcomings and limitations such as intensive preparation processes or assisting materials with special properties. Also, most of these techniques lack precise control over the positioning and orientation of individual SWNTs. Therefore, they may not be appropriate for scaled
 up fabrication of individual SWCNTs devices.
 
 Dielectrophoresis technique first adopted by Pohl \cite{pohl1951motion} is the motion of suspended particles relative to that of the solvent resulting from polarization forces produced by an inhomogeneous electric field. This technique is simple but versatile and can be conducted at room temperature with low voltage so it can be used to align CNTs between electrodes with high yield \cite{vijayaraghavan2007ultra,vijayaraghavan2007ultra,stokes2010high}. Several parameters have great importance
 in dielectrophoresis technique such as alternating voltage (AC) amplitude, frequency of the applied voltage, deposition
 time, the geometry of electrodes and solution concentration.

\new{In \cite{an2013dielectrophoretic}, multi-walled carbon nanotubes with a variety of sizes were assembled onto electrodes using alternating current electric fields by dielectrophoresis. Assembling CNTs between electrodes using a combination of optically induced dielectrophoresis force and dielectrophoresis force was performed in \cite{wu2011assembly}. A different study for the directed assembly of CNTs and graphene in an alternating current (AC) electric field can be found in \cite{vijayaraghavan2013bottom}. The authors of \cite{abdulhameed2019role} used simulation results 
to clarify the role of the medium on the motion of CNTs during alignment using a DEP process prior to actual alignment. 
	}

  One of the most important problems in using this technique for aligning CNTs between electrodes is that metallic carbon
  nanotubes do always exist along with semiconducting CNTs in the solution so adjusting appropriate parameters to absorb only semiconductor CNTs is a great consideration. 
  
  In this paper, we focus on the frequency of the applied voltage and show that with a suitable theoretical model for this technique we can find the right frequency (one of the important parameters in the process) in this process to absorb and align semiconductor CNTs between electrodes and consequently a p-type CNT-FET behavior of the fabricated device. 
  
  Here we first derive and demonstrate a prevalent model for DEP,
  namely a model based on a chain of spherical particles along the CNT axis (\cite{xi2011nano}, pp. 29-49) and then use the new model,
  a prolate ellipsoid, for CNTs. Using the new model, the frequency of the applied voltage can be chosen properly as different values are calculated compared to the previous model, to absorb semiconductor CNTs. This new theoretical result draws out the reason devices show metallic or low bandgap semiconductor behavior, as our experiment suggested. On the other hand, it presents pertinent values in terms of frequency to yield semiconducting characteristics. The calculated results
  are consistent with experiments, yet previously calculated parameters in a particular frequency have led
  to mostly metallic or quasi-metallic behavior. \new{The theoretical results can be extended to solve different novel problems in electrochemistry, e.g., \cite{abbaszade2012fourth,abbaszadeh2019error,dehghan2019error,dehghan2018efficient},  modern physics \cite{dehghan2018efficient,dehghan2006finite,dehghan2016legendre,khodadadian2019multilevel,dehghan2005solution,mohammadi2019numerical} as well as simulation of silicon nanowire sensors \cite{khodadadian2016basis,mirsian2019new,khodadadian2017optimal,khodadadian2018three,khodadadian2020bayesian}, field-effect transistors \cite{khodadadian2018optimal,khodadadian2020adaptive}, and ion channels \cite{khodadadian2015transport}.}

The paper is organized as follows:  In Section 2, we explain the theoretical work and obtain the frequency interval for the spherical model. In Section 3, we describe the ellipsoid model and estimate the frequency interval related to the model. In Section 4, we employ the obtained frequencies (according to spherical and ellipsoid models) and in order to identify the type (metallic or semiconductor) of aligned SWCNTs, we plot I-V characteristics of fabricated devices. Finally, the conclusions are explained in Section 5.

\section{The theoretical work}
The basic idea in the theory of dielectrophoresis is that a dielectric particle in a nonuniform electric field experiences a
net force which depends on the electric field and effective dipole moment of the particle generated because of the external
electric field. 


Dielectrophoresis is a phenomenon in which a force is exerted on a dielectric particle when it is suspended in a non-uniform electric field. The polarization of particles by non-uniform electric field makes it possible to exert DEP force even on non-charged particles. All polarizable particles exhibit Dielectrophoresis phenomena in the presence of an electric field. Electrical properties of particle and medium, shape and size of the particle, the strength of the electric field and frequency of applied field are the main parameters that determine the strength of the exerted force. 

The dipole consists of equal and opposite charges ($+q$) and ($-q$) located a distance vector  \new{$\boldsymbol{d}$} apart, and it is located in an
electric field  \new{$\boldsymbol{E}$}.  
If the electric field is nonuniform, then, in general, the two charges ($+q$ and $-q$) will experience different values of the
Vector field  \new{$\boldsymbol{E}$}  and the dipole will experience a net force. The total force on the dipole is
\begin{align}
 \new{\boldsymbol{F}} =q\new{\boldsymbol{E}}(\new{\boldsymbol{t}}+\new{\boldsymbol{d}})-q\new{\boldsymbol{E}}(\new{\boldsymbol{r}}),
\end{align}
where \new{$\boldsymbol{r}$} is the position of the $-q$ particle. The first order approximation (\textit{dielectrophoretic approximation}) is given by
\begin{align}\label{4}
\new{\boldsymbol{F}_{\text{dipole}}}=(q\new{\boldsymbol{d}}\cdot \nabla)\new{\boldsymbol{E}}(\new{\boldsymbol{r}}).
\end{align}
So given the electric field to calculate the force on a dipole we should first calculate the effective moment of the dielectric
particle. Here we are trying to model SWCNTs suspended in a medium. Our first model which is prevalent is the spherical particle model of CNTs ( a chain of spherical particles) so we bring the main
formula regarding effective dipole moment of a spherical dielectric particle in an external electric field and then go to a
new model.

The problem of a spherical dielectric particle in an external electric field can (see \cite{nayfeh2015electricity}, pages 149-152). Considering
the electric field in the direction $\hat{z}$ and solving Laplace’s equation we have for potential functions inside and outside of the sphere:
\begin{align}
\phi_1(r,\theta)&=-E_0 r\cos \theta + \left(\frac{\varepsilon_p -\varepsilon_m}{\varepsilon_p+2\varepsilon_m} R^3 E_0 \right) \frac{\cos \theta}{r^2},\qquad &&r\geq R,\label{5}\\
\phi_2(r,\theta)&=-\left(\frac{3\varepsilon_m}{\varepsilon_p+2\varepsilon_m} E_0\right) r \cos\theta, \qquad &&r < R,\label{6}
\end{align} 
 where $\varepsilon_m$ is the permittivity of the medium and $\varepsilon_p$ that of particle \new{and the second term indicates the dipole}. Furthermore, we should note that the electrostatic potential $\phi_{\text{dipole}}$ dipole due to a point dipole of moment $\rho_{\text{eff}}$ in a dielectric medium of
 permittivity $\varepsilon_m$ is (see \cite{nayfeh2015electricity}, page 50)
\begin{align}
\label{7}
\phi_{\text{dipole}}=\frac{\rho_{\text{eff}} \cos \theta}{4 \pi \varepsilon_m r^2}.
\end{align}
By comparing \eqref{7} and \eqref{5} we conclude
\begin{align}
\rho_{\text{eff}}=4\pi\varepsilon_m \frac{\varepsilon_p-\varepsilon_m}{\varepsilon_p+2\varepsilon_m}R^3 E_0
\end{align}
consequently for the special case of the homogeneous, dielectric sphere, the expression for the effective dipole moment
is
\begin{align}
\rho_{\text{eff}}=4\pi \varepsilon_m \mathcal{K} R^3 E_0.
\end{align}
Here, $\mathcal{K}$ known as the Clausius-Mossotti function \cite{pethig2017dielectrophoresis} provides a measure of the strength of the effective polarization of a spherical particle as a function of $\varepsilon_p$ and $\varepsilon_m$
\begin{align}
\mathcal{K} (\varepsilon_p,\varepsilon_m)=\frac{\varepsilon_p-\varepsilon_m}{\varepsilon_p+2\varepsilon_m}.
\end{align}
When $\varepsilon_p>\varepsilon_m$ then $\mathcal{K}$ and the effective dipole moment $\rho_{\text{eff}}$ is collinear with the electric field vector and when $\varepsilon_p<\varepsilon_m$ two vectors are antiparallel.

Now we consider the electric field as an AC electric field and assume that sphere and medium have a finite conductivities.
\begin{align}
\new{\boldsymbol{E}}(t)=E_0\exp (i\omega t)\, \hat{z}.
\end{align}
The solution is the same as (\ref{5}) and (\ref{6}) except for coefficients that now become complex and we arrive at a more general expression for the complex effective moment
\begin{align}\label{13}
\rho_{\text{eff}}=4\pi\varepsilon_m\bar{K}R^3\new{\boldsymbol{E}}(t).
\end{align}
The Clausius-Mossotti factor now has become a function of the complex permittivities, containing magnitude
and phase information about the effective dipole moment. It considers the complex coordinates as 
\begin{align}
\bar{\varepsilon_m}=\varepsilon_m-i\frac{\sigma_m}{\omega}\qquad \bar{\varepsilon_p}=\varepsilon_p-i\frac{\sigma_p}{\omega}
\end{align}
where $\sigma_m$ and $\sigma_p$ indicate respectively the conductivity of medium and the particle. Such ohmic, dispersive behavior is a consequence of the finite time required to build up surface charge $\sigma$ at the interface.

Assuming spatial variation for electric field from equations (\ref{13}) and (\ref{4}) we have for time-averaged force
\begin{align}\label{15}
\new{\boldsymbol{F}}_{\text{DEP}}=\frac{1}{2}\mathcal{R}\left\{ (\rho_{\text{eff}}\cdot\nabla) \new{\boldsymbol{\bar{E}}}(t,x) \right\}=\pi \varepsilon_m \mathcal{R}\{ K\} R^3 \nabla (E^2) 
\end{align}
where $\new{\boldsymbol{\bar{E}}}(t,x)$ indicates the complex conjugate of $ \new{\boldsymbol{E}}(t,x)$ and the real and imaginary parts are given by  \cite{pethig2017dielectrophoresis}
\begin{align}
\mathcal{R} \{\K (\bar{\varepsilon_p},\bar{\varepsilon_m}) \}&=\frac{(\varepsilon_p-\varepsilon_m)(\varepsilon_p+2\varepsilon_m) + (\frac{1}{\omega})^2 (\sigma_p-\sigma_m)(\sigma_p+2\sigma_m)}{(\varepsilon_p+2\varepsilon_m)^2+(\frac{1}{\omega})^2 (\sigma_p+2\sigma_m)^2} \\
\mathcal{I} \{\K (\bar{\varepsilon_p},\bar{\varepsilon_m})&=\frac{3}{\omega} \frac{\varepsilon_p \sigma_m -\varepsilon_m \sigma_p}{(\varepsilon_p+2\varepsilon_m)^2 + \frac{1}{\omega})^2 (\sigma_p+2\sigma_m)^2}.
\end{align}
Also, the relevant relaxation time constant in this case is $\tau=\frac{\varepsilon_p+2\varepsilon_m}{\sigma_p+2\sigma_3}$ is the reciprocal of the frequency for which $\mathcal{R} \{\K \}$ is maximum. The important point about this constant
is that whenever relaxation time constant is bigger than the reciprocal of applied frequency, the particle will respond to
time-averaged force and torque, i.e., we have $\tau>\frac{1}{\omega_{\text{applied~voltage}}}$.

 From (\ref{15}) it is clear that to absorb CNTs to the region of the strong electric field (downward between electrodes)
the term $\mathcal{R} \{\K\}$ should be positive. Assuming $\varepsilon_p\leq \varepsilon_m$ and $\sigma_p\geq \sigma_m$ (large energy gap semiconductor SWCNTs in
aqueous solution) we have the following formula for the zero force-frequency (frequency for which particle experiences
no time-averaged force)
\begin{align}\label{20}
\omega\left(\mathcal{R} \{\K (\bar{\varepsilon_p},\bar{\varepsilon_m})\}=0\right)=\sqrt{\frac{(\sigma_p-\sigma_m)(\sigma_p+2\sigma_m)}{|(\varepsilon_p-\varepsilon_m)(\varepsilon_p+2\varepsilon_m)|}}.
\end{align}
The key point is obtaining the below frequency that we desire from the direction of force to align nanotubes between
electrodes.

\begin{figure}[t!]
	\centering
	\includegraphics[width=0.7\linewidth]{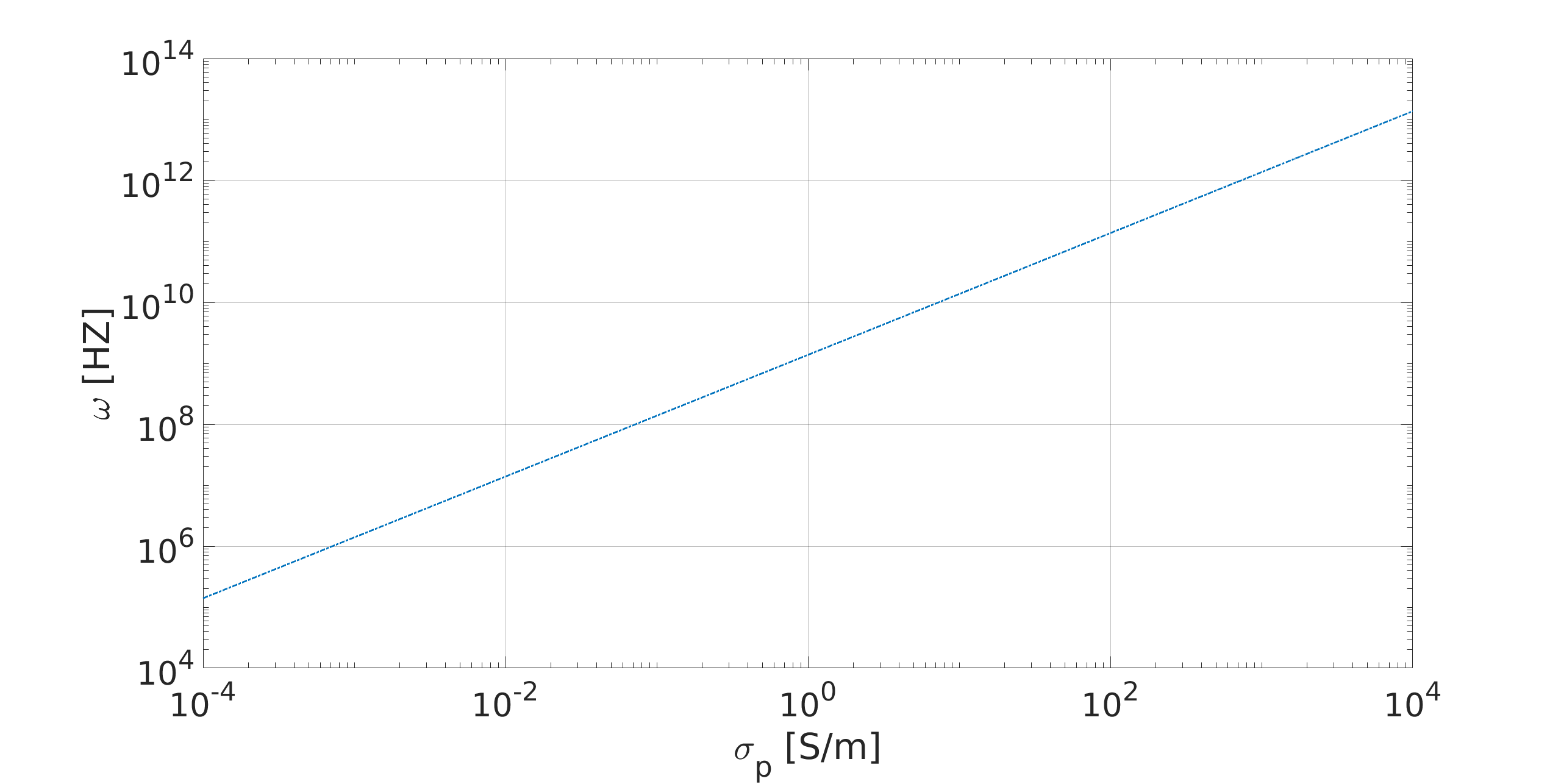}
	\caption{The frequency as a function of $\sigma_{\text{p}}$.}
	\label{fig3}
\end{figure}

Considering SWCNTs solved in deionized water, we use Eq. \ref{20} to calculate the frequency of zero force where $\varepsilon_p=30\,\varepsilon_0$, $\sigma_p=\unit{10^{-2}}{S/m}$, $\varepsilon_m=70\,\varepsilon_0$, and  $\sigma_m=\unit{5.5\times10^{-6}}{S/m}$ are assumed. Here, we have $\varepsilon_p\approx \varepsilon_0$ for large gap semiconductor SWNTs and larger values for small gap and metallic tubes. Therefore, we conclude
\begin{align}
\omega\left(\mathcal{R} \{\K (\bar{\varepsilon_p},\bar{\varepsilon_m})\}=0\right)\cong\unit{13.7}{MHz}\qquad\frac{1}{\tau}\cong \unit{6.6}{MHz}
\end{align}
So to attract an SWCNT with a given parameters frequency of the applied field should be in the following interval
\begin{align}
\unit{6.6}{MHz} < \omega<\unit{13.7}{MHz}
\end{align}
An important characteristic of equation (\ref{20}) is that $\omega\left(\mathcal{R} \{\K (\bar{\varepsilon_p},\bar{\varepsilon_m})\}=0\right)$ increases with $\sigma_p$
\begin{align}
\frac{\partial}{\partial \sigma_p}\omega^2\left(\mathcal{R} \{\K (\bar{\varepsilon_p},\bar{\varepsilon_m})\}=0\right)=\frac{(2\sigma_p+\sigma_m)}{|(\varepsilon_p-\varepsilon_m)(\varepsilon_p+2\varepsilon_m)|}>0
\label{24}
\end{align}
This means that the frequency of zero force shifts to the right with the increment in the conductivity of particle so to absorb SWCNTs with lower conductivity (wide-bandgap) we should go to lower frequencies and vice versa. Figure \ref{fig3} shows the dependency of the frequency to $\sigma_{\text{p}}$ varying between $\unit{10^{-4}}{S/m}$ and $\unit{10^{4}}{S/m}$.

\section{Ellipsoid particle model of CNTs}
We now use a more realistic model namely, prolate ellipsoid, to find the right interval for the frequency of applied voltage.
\new{In \cite{gimsa2001comprehensive}, an ellipsoid particle model in an electric field was introduced.
The model predicts the potential at the ellipsoid's surface leading to the induced dipole moment. In this section, we strive to introduce the ellipsoid model of nanotubes to find proper frequency intervals of the applied electric field to absorb semiconductor nanotubes and compare the result to the prevalent model (sphere model). In other words, the main goal is to show that the ellipsoid model is better suited to model cylindrical shape nanotubes, and by using this model we can find a better estimate of the frequency interval of the applied electric field to absorb semiconductor carbon nanotubes.
}

The problem of the ellipsoid in an electric field can be solved by introducing elliptical coordinates \cite{landau2013electrodynamics}.
We should note that in the case of an SWCNT, we have different permittivities and conductivities in different
directions. In the case of semiconductors, SWCNT's longitudinal polarizability factor typically is more than ten times
higher than transverse polarizability \cite{kozinsky2006static}.  Furthermore, these dipoles generated in the middle part cancel each other pairwise
so we have a high aspect ratio, i.e., a high length to radios ratio for the generated dipole.

\begin{figure}[t!]
	\centering
	\includegraphics[width=1\linewidth]{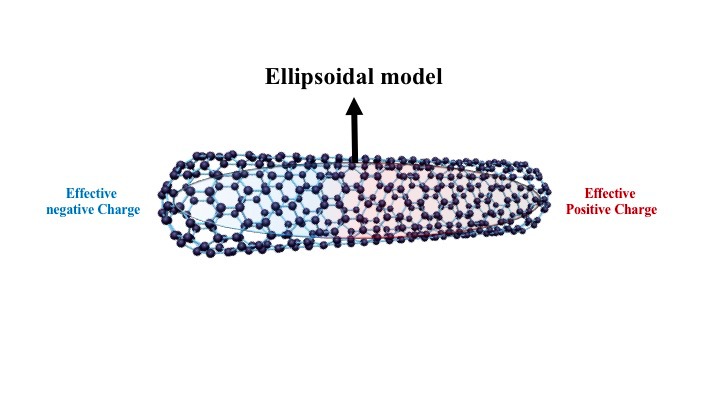}
	\vspace{-3.5cm}
	\caption{Polarization of a SWCNT.}
	\label{fig1}
\end{figure}

Figure \ref{fig1} illustrates the general idea of the ellipsoid model for the SWCNT dipole. In an external electric field, all atoms get polarized and those atoms located near each other overlap with each other and we get two main centers for two opposite
charge the positive one in the leftmost side and a negative one in the rightmost side so we have a dipole with a characteristic
length as the length of the SWCNTs.

The resulting equation for the induced dipole moment in the direction $n$ is
\begin{align}\label{25}
\rho_{\text{eff}-n}=\frac{4 \pi a_1 a_2 a_3}{3} \varepsilon_m \left[ \frac{\bar{\varepsilon}_{\text{p-n}} -\bar{\varepsilon}_m}{\bar{\varepsilon}_m +A_n (\bar{\varepsilon}_{\text{p-n}} -\bar{\varepsilon}_m)} \right]E_n,
\end{align}
where $A_n$ is depolarization factor for the axis $n$
\begin{align}\label{26}
A_n=\frac{a_1a_2a_3}{2}\int_{0}^{\infty} \frac{dx}{(x+a_n^2)\sqrt{(x+a_1^2)(x+a_2^2)(x+a_3^2)}},
\end{align}
Where $a_1$, $a_2$, and $a_3$ are half length along the axes and $\bar{\varepsilon}_{\text{p-n}}$ is the complex permittivity in the direction $n$. Comparing equation (\ref{13}) to (\ref{25}), we conclude that in this case for Clausius-Mossotti factor we have
\begin{align}
\K_n(\bar{\varepsilon}_{\text{p}},\bar{\varepsilon}_{\text{m}},a_1,a_2,a_3)=  \frac{1}{3}\frac{\bar{\varepsilon}_{\text{p-n}} -\bar{\varepsilon}_m}{\bar{\varepsilon}_m +A_n (\bar{\varepsilon}_{\text{p-n}} -\bar{\varepsilon}_m)} 
\end{align}
If $a_1 = a_2 = a_3$, all three $A_n$s are $\frac{1}{3}$
so we get the same result for Clausius-Mossotti factor as before for
sphere. If $a_1\gg a_2=a_3$ that we are concerned, $A_n$s can be calculated as follows
\begin{align}\label{28}
A_1=\frac{a_1a_2a_3}{2}\int_{0}^{\infty} \frac{dx}{(x+a_2^2)(x+a_1^2)^{\frac{3}{2}}}=\frac{1}{2}\frac{1-e^2}{e^3}\left[ \ln \frac{1+e}{1-e}-2e\right],
\end{align}
where $e$ is the eccentricity and is defined as follows
\begin{align}
e=\sqrt{1-\left(\frac{a_2}{a_1}\right)^2}\sim 1-\frac{1}{2} \left(\frac{a_2}{a_1}\right)^2+\frac{1}{8} \left(\frac{a_2}{a_1}\right)^4+\cdots.
\end{align}
For very long needle-shaped ellipsoid we have following approximation for $A_1$
\begin{align}
A_1\sim \left(\frac{a_2}{a_1}\right)^2 \left(\ln \frac{2a_1}{a_2}-1 \right).
\end{align}
For $A_2$ and $A_3$ we get
\begin{align}\label{30}
A_2=A_3=\frac{a_1a_2^2}{2}\int_{0}^{\infty} \frac{dx}{(x+a_2^2)\sqrt{(x+a_1^2)}}=\frac{1}{2e^2}-\frac{1}{2}\frac{1-e^2}{4e^3}\left[ \ln \frac{1+e}{1-e} \right].
\end{align}
In the case of a SWCNT with diameter to length ratio of $\frac{a_2}{a_1}$ (which is relevant in our case) we have $A_1\sim 6.6\times10^{-6}$ and $A_2=A_3\sim \frac{1}{2}$. From equation \eqref{25} and (\ref{4}) we get the following formula for the time averaged force exerted on a SWCNT in this model
\begin{align}\label{32}
\new{\boldsymbol{F}}_{\text{ellipsoidal~dipole}}=\frac{2\pi a_1 a_2 a_3}{3}\varepsilon_m\left(\mathcal{R}\left\{ \frac{\bar{\varepsilon}_{\text{p-n}} -\bar{\varepsilon}_m}{\bar{\varepsilon}_m +A_n (\bar{\varepsilon}_{\text{p-n}} -\bar{\varepsilon}_m)} E_n\frac{\partial}{\partial x_n} \right\}  \right)\new{\boldsymbol{E}}
\end{align}
As before we need real and imaginary parts of Clausius-Mossotti factor which in this case from equation (\ref{32}) we get
\begin{align}
\mathcal{R}\left(\K_n(\bar{\varepsilon}_{\text{p}},\bar{\varepsilon}_{\text{m}},a_1,a_2,a_3)\right)&=\frac{1}{3} \frac{(\varepsilon_{\text{p-n}}-\varepsilon_{\text{m}}) \left(A_n \varepsilon_{\text{p-n}} +(1-A_n)\varepsilon_m\right)+\left(\frac{1}{\omega}\right)^2(\sigma_{\text{p-n}}-\sigma_{\text{m}}) \left(A_n \sigma_{\text{p-n}} +(1-A_n)\sigma_m\right)}{\left(A_n \varepsilon_{\text{p-n}} +(1-A_n)\varepsilon_m\right)^2+\left(\frac{1}{\omega}\right)^2  \left(A_n \sigma_{\text{p-n}} +(1-A_n)\sigma_m\right)^2}\\
\mathcal{I}\left(\K_n(\bar{\varepsilon}_{\text{p}},\bar{\varepsilon}_{\text{m}},a_1,a_2,a_3)\right)&=
\frac{1}{3\,\omega} \frac{\varepsilon_{\text{p-n}}\sigma_{\text{m}} -\varepsilon_m\sigma_{\text{p-n}}}{\left(A_n \varepsilon_{\text{p-n}} +(1-A_n)\varepsilon_m\right)^2+\left(\frac{1}{\omega}\right)^2  \left(A_n \sigma_{\text{p-n}} +(1-A_n)\sigma_m\right)^2}
\end{align}
The relaxation time constant for each directions in this case is
\begin{align}
\tau_n=\frac{A_n\varepsilon_{\text{p-n}}+(1-A_n)\varepsilon_{\text{m}}}{A_n\sigma_{\text{p-n}} +(1-A_n) \sigma_m}
\end{align}
The point here is that an ellipsoid particle in an electric field always aligns itself with its longest axes along the direction
of electric field so in this case, the ruling term is $\tau_1$.

In the case of a long SWCNT with a high aspect ratio, we get the following approximate formula for the experienced force by
SWCNT
\begin{align}\label{36}
\new{\boldsymbol{F}}_{\text{ellipsoidal~dipole}}=\frac{\pi a_1 a_2 a_3}{3}\varepsilon_m\mathcal{R}\left\{ \frac{\bar{\varepsilon}_{\text{pl}} -\bar{\varepsilon}_m}{\bar{\varepsilon}_m +A_1 (\bar{\varepsilon}_{\text{pl}} -\bar{\varepsilon}_m)}  \right\}  \nabla E^2
\end{align}
where $\bar{\varepsilon}_{\text{pl}}$ is the longitudinal complex permittivity. Frequency dependence of the force is present in the $\mathcal{R}\left\{ \frac{\bar{\varepsilon}_{\text{pl}} -\bar{\varepsilon}_m}{\bar{\varepsilon}_m +A_1 (\bar{\varepsilon}_{\text{pl}} -\bar{\varepsilon}_m)}  \right\}$ term. With this approximation we have following range of the applied frequency:
\begin{align}
\frac{1}{\tau_1}<\omega<\omega\left(\mathcal{R}\left(\K_1(\bar{\varepsilon}_{\text{pl}},\bar{\varepsilon}_{\text{m}},a_1,a_2,a_3)=0\right) \right)
\end{align}
where the right hand side indicates
\begin{align}
\omega\left(\mathcal{R}\left(\K_1(\bar{\varepsilon}_{\text{pl}},\bar{\varepsilon}_{\text{m}},a_1,a_2,a_3)=0\right) \right)=\sqrt{\frac{(\sigma_{\text{pl}}-\sigma_{\text{m}}) (A_1\sigma_{\text{pl}}+ (1-A_1) \sigma_{\text{m}})}{(\varepsilon_{\text{pl}}-\varepsilon_{\text{m}}) (A_1\varepsilon_{\text{pl}}+ (1-A_1) \varepsilon_{\text{m}})}}
\end{align}

\begin{table}[h!]
	\centering
	\new{\begin{tabular}{*{9}l}
		parameter & $\varepsilon_{\text{pl}}$&  $\varepsilon_{\text{pt}}$ & $\sigma_{\text{pl}}$ &  $\sigma_{\text{pt}}$ & $\varepsilon_{\text{m}}$& $\sigma_{\text{m}}$ & $A_1$  \\
		\hline
		value & 30 $\varepsilon_0$ & $8\varepsilon_0$ & $\unit{10^{-2}}{S/m}$ & 0& $70\varepsilon_0$ & $\unit{5.5\times10^{-6}}{S/m}$ & $6.6\times10^{-6}$   \\
	\end{tabular}}
	\caption{The parameter values used to obtain the frequency.}
	\label{table2}
\end{table}

\begin{figure}[t!]
	\centering
	\includegraphics[width=0.6\linewidth]{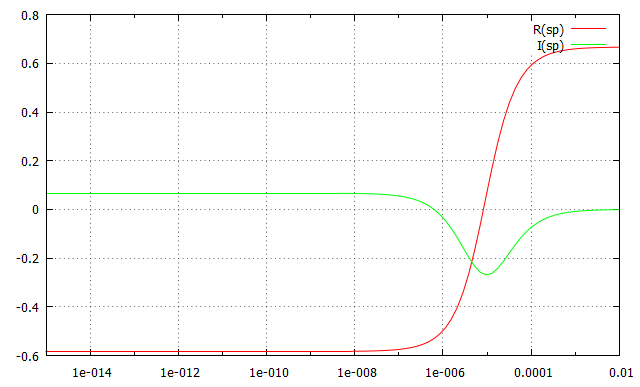}
	\caption{Real and imaginary parts of CM factor against transverse conductivity in $\unit{10}{KHz}$ frequency}
	\label{fig7}
\end{figure}

In order to estimate the frequency, we use the efficient parameters, summarized in Table \ref{table2}. Accordingly we get
\begin{align}
\unit{9}{KHz}<\omega<\unit{500}{KHz}
\end{align}

Here like Eq. \ref{24}, with increasing electrical conductivity of SWCNTs, frequency of zero force shift to the right and also relaxation time constant increases so to absorb SWCNTs with higher electrical conductivity we should use higher frequencies which results in absorption of more metal SWCNTs and as a result fabricated device shows metallic behavior. besides that, because of the presence of other forces that result in different motions like electro thermal motion or Brownian motion we should choose the frequency near the lower limit to have higher DEP force and near the stable polarization. 

An important point here is that the transverse conductivity of SWCNT is much lower than its longitudinal conductivity and
consequently we assume it to be zero. Figure \ref{fig7} shows that variation in this assumption has not a considerable effect on the real and imaginary part of the Clausius-Mossotti factor. According to Figure \ref{fig7} variation in $\mathcal{R}$ is restricted to the interval $(-0.6, 0.7)$ and for I restricted to the interval $(-0.3, 0.1)$ so
totally $\mathcal{R}$ is restricted to $(-1.3, 1.2)$ and $\mathcal{I}$ to $(-0.3, 0.2)$ which are not comparable to longitudinal factors at the relevant
frequency.

The point here is that, because of the presence of other forces which result in different motions like electrothermal
motion or Brownian motion we should choose the frequency near the lower limit to have higher DEP force and near the
stable polarization.

\section{Experimental results}

In order to identify the type (metallic or semiconductor) of aligned SWCNTs with two different applied frequencies, based on the two models, we plot I-V characteristics of fabricated devices. We know that for metallic SWCNTs we would observe a straight line in the I-V plane and for semiconductor SWCNTs we expect a nonlinear relationship between current and applied voltage. Indeed, for fabricated devices with a dominant number of semiconductor SWCNTs, with the increment in voltage more and more holes participate in the current generation and we would see an exponential I-V characteristic.

The SWCNTs used in our experiment are in the form of aqueous solution and purchased from Nanointegris
\href{http://www.nanointegris.com/}{(http://www.nanointegris.com/)} specification for the SWNTs solution are: diameter range from $\unit{1.2}{nm}$ to $\unit{1.7}{nm}$, length range varies between
$\unit{300}{nm}$ and 5 microns, metal catalyst impurity < 1\%, amorphous carbon impurity 1-5\%, electronic enrichment 98\%
semiconductor SWCNT. We diluted the solution to $\unit{100}{ng/ml}$ and used in the DEP process.

\begin{figure}[ht!]
	\centering
	\subfloat{\includegraphics[width=0.45\textwidth]{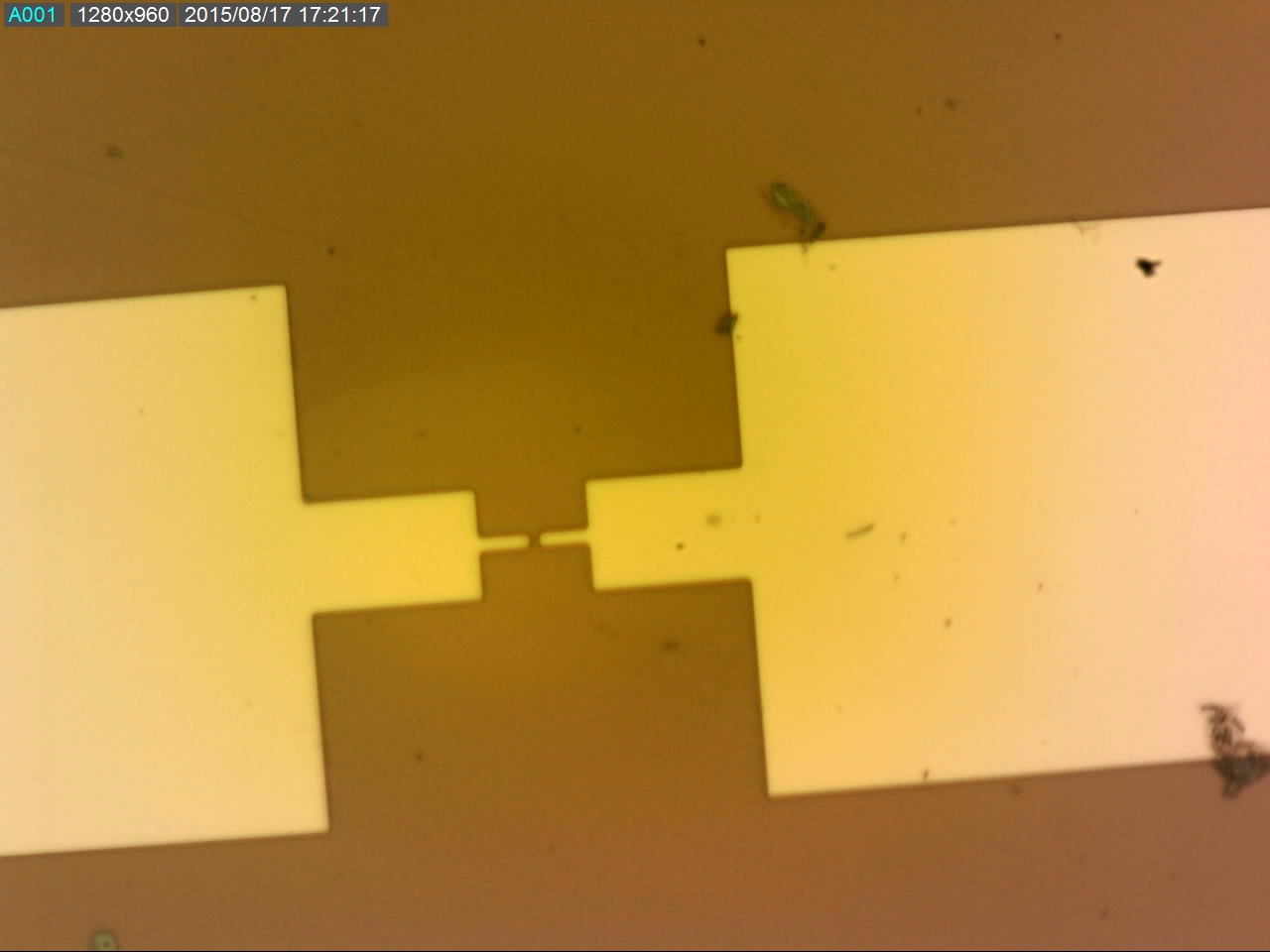}}
	\hfill
	\subfloat{\includegraphics[width=0.45\textwidth]{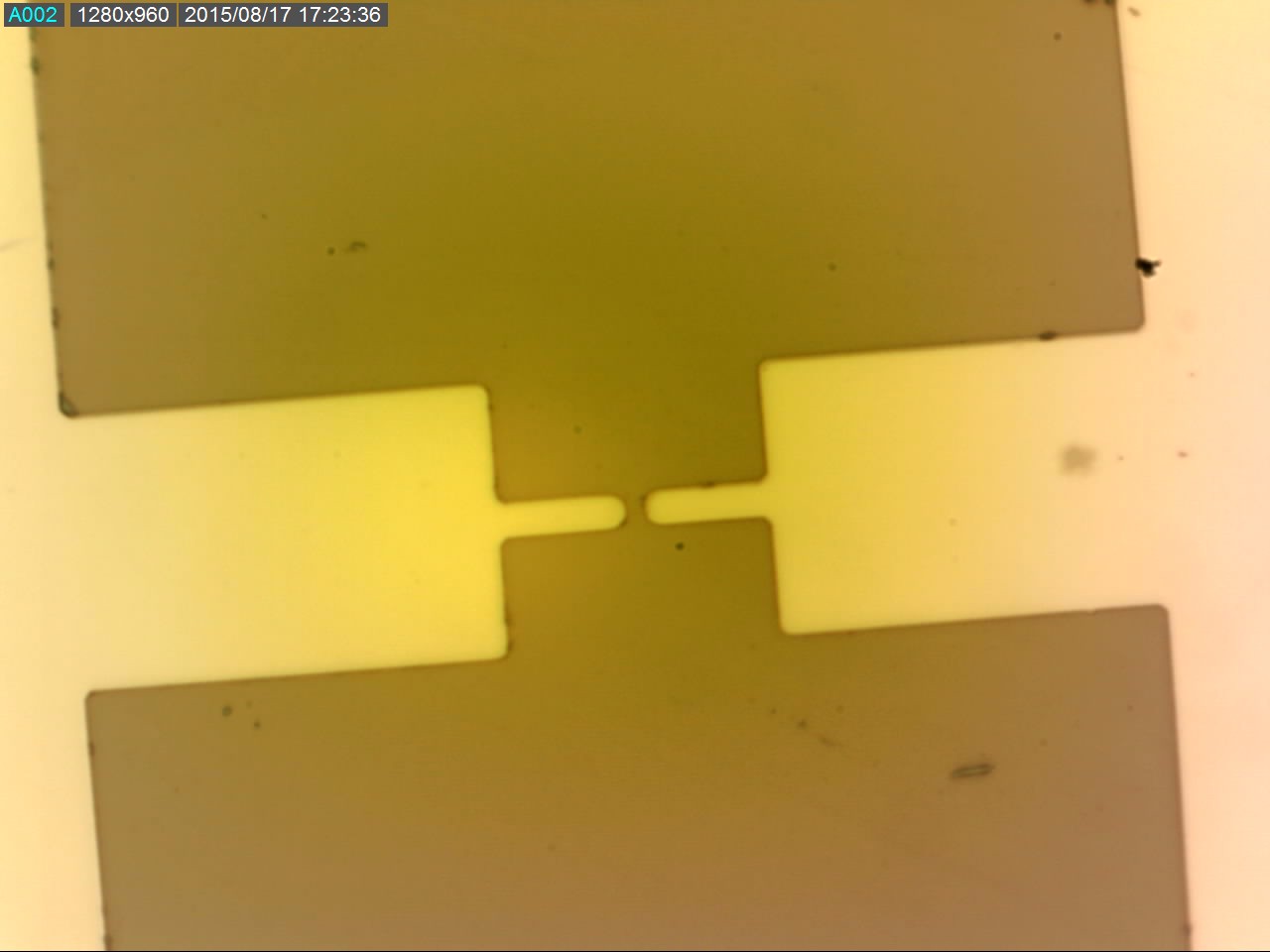}}
	\newline\\\vspace{0.2cm}
	\subfloat{\includegraphics[width=0.45\textwidth]{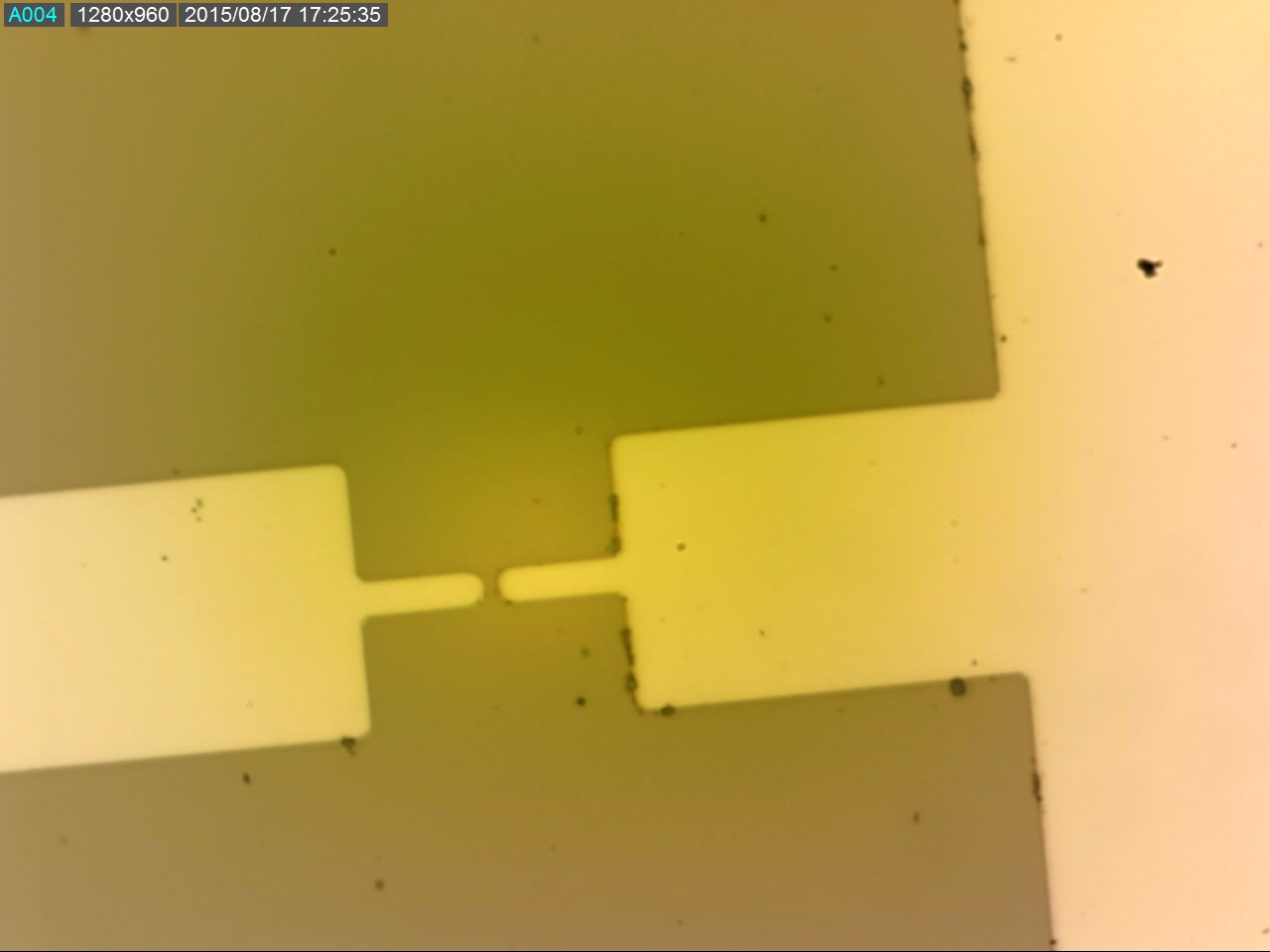}}
	\hfill
	\subfloat{\includegraphics[width=0.45\textwidth]{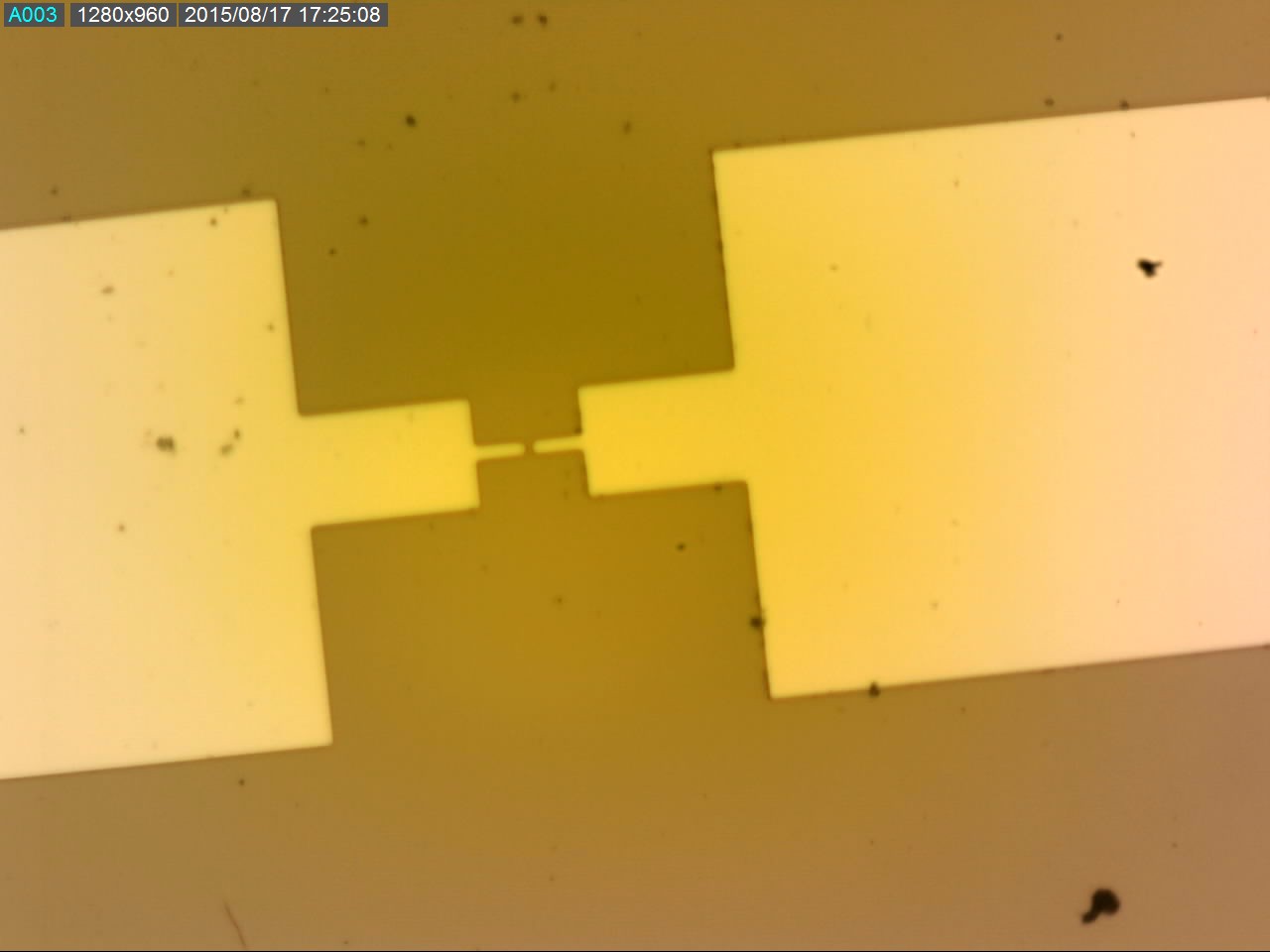}}
	\caption{Electrodes with 2 microns gap (top) and 1 microns gap (bottom).}
	\label{fig:electrodes}
\end{figure}

For the source and drain electrodes, we used two kinds of electrodes one with 2-microns gap and another with a 1-micron gap between electrodes both with 3-microns width at the end tip. The electrodes are shown in Figure \ref{fig:electrodes}.

We performed the DEP process and in order to align the SWCNTs  between the electrodes an applied voltage of 
$\unit{6}{V}$ amplitude for 2 microns gap electrodes and $\unit{5}{V}$ for the 1-micron gap is applied.

 \new{Now we strive to use the obtained theoretical results (estimated frequencies) to align the SWCNTs. The prevalent model (sphere model) suggests that to absorb semiconductor carbon nanotubes with given parameters, we should use an AC electric field with a frequency near $\unit{7}{MHz}$. The second model (ellipsoid) suggests using frequency as low as $\unit{10}{kHz}$. In the experiment we will use $\unit{50}{kHz}$.
	}

\begin{figure}[t!]
	\centering
	\includegraphics[width=10cm,height=5.5cm]{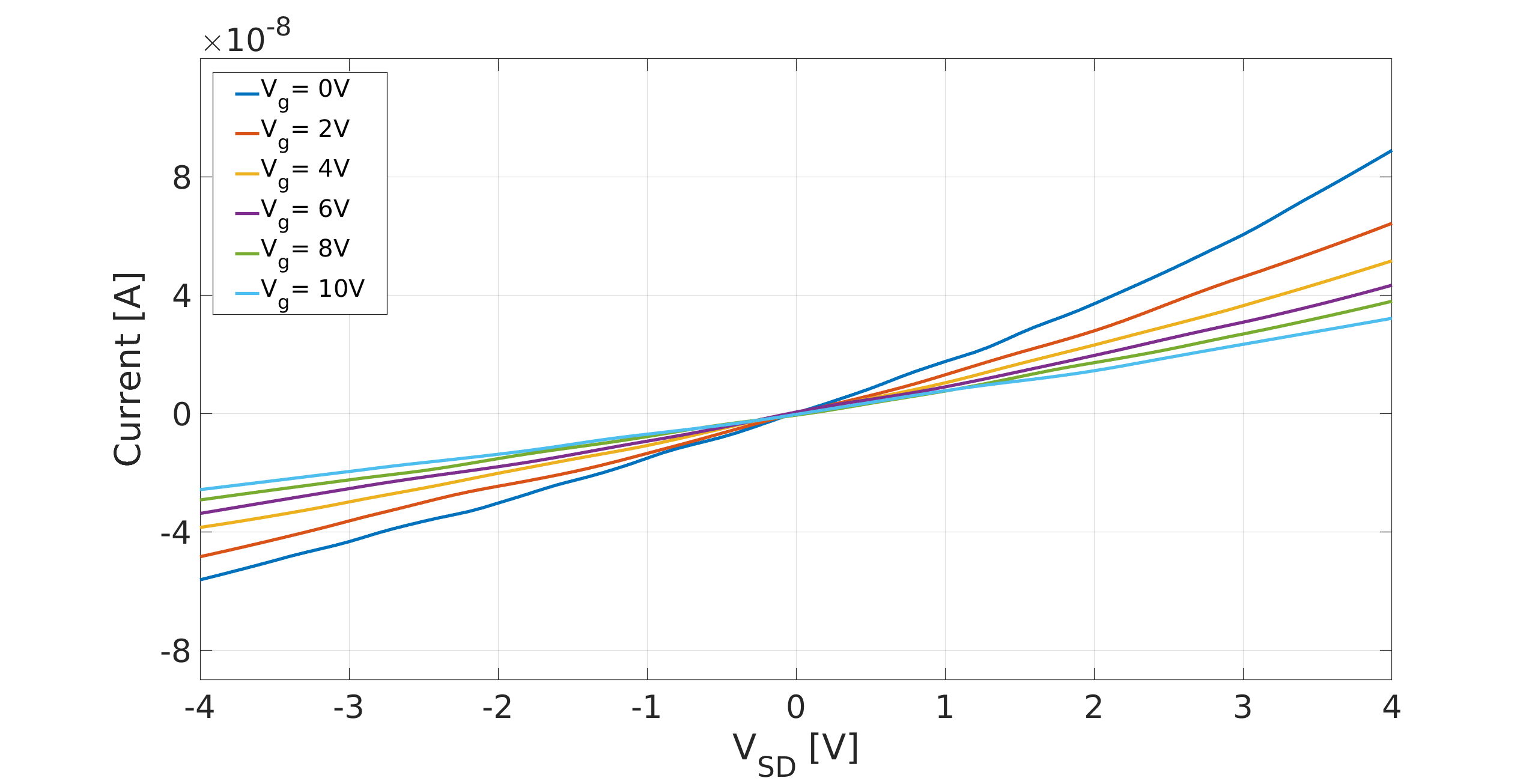}
	\caption{I-V characteristics of fabricated device with 1 micron gap electrodes with $\unit{7}{MHz}$ frequency.}
	\label{fig10}
\end{figure}

\begin{figure}[t!]
	\centering
	\includegraphics[width=10cm,height=5.5cm]{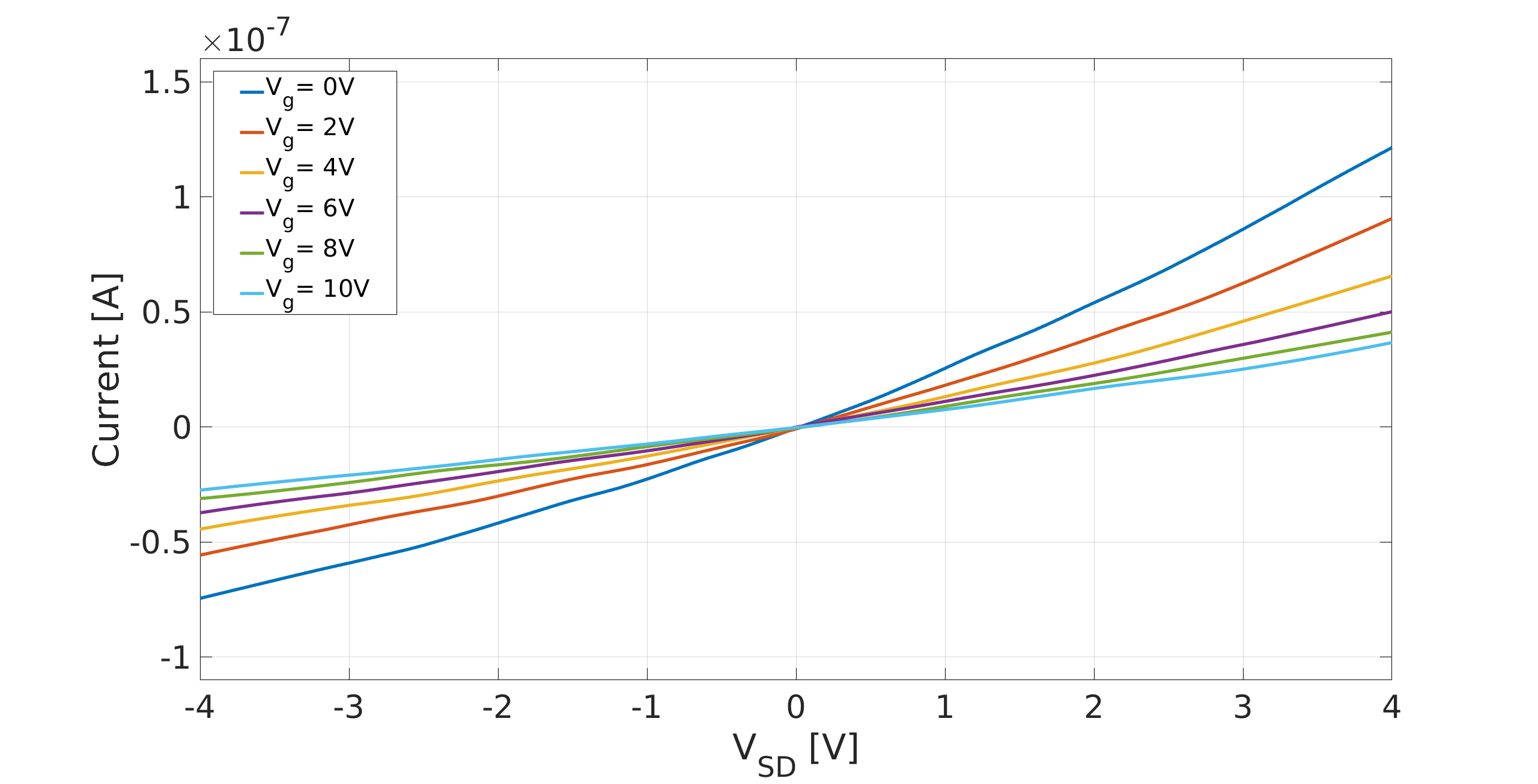}
	\caption{I-V characteristics of fabricated device with 2 micron gap electrodes with $\unit{7}{MHz}$ frequency.}
	\label{fig11}
\end{figure}

\begin{figure}[t!]
 	\centering
 	\includegraphics[width=10cm,height=5.5cm]{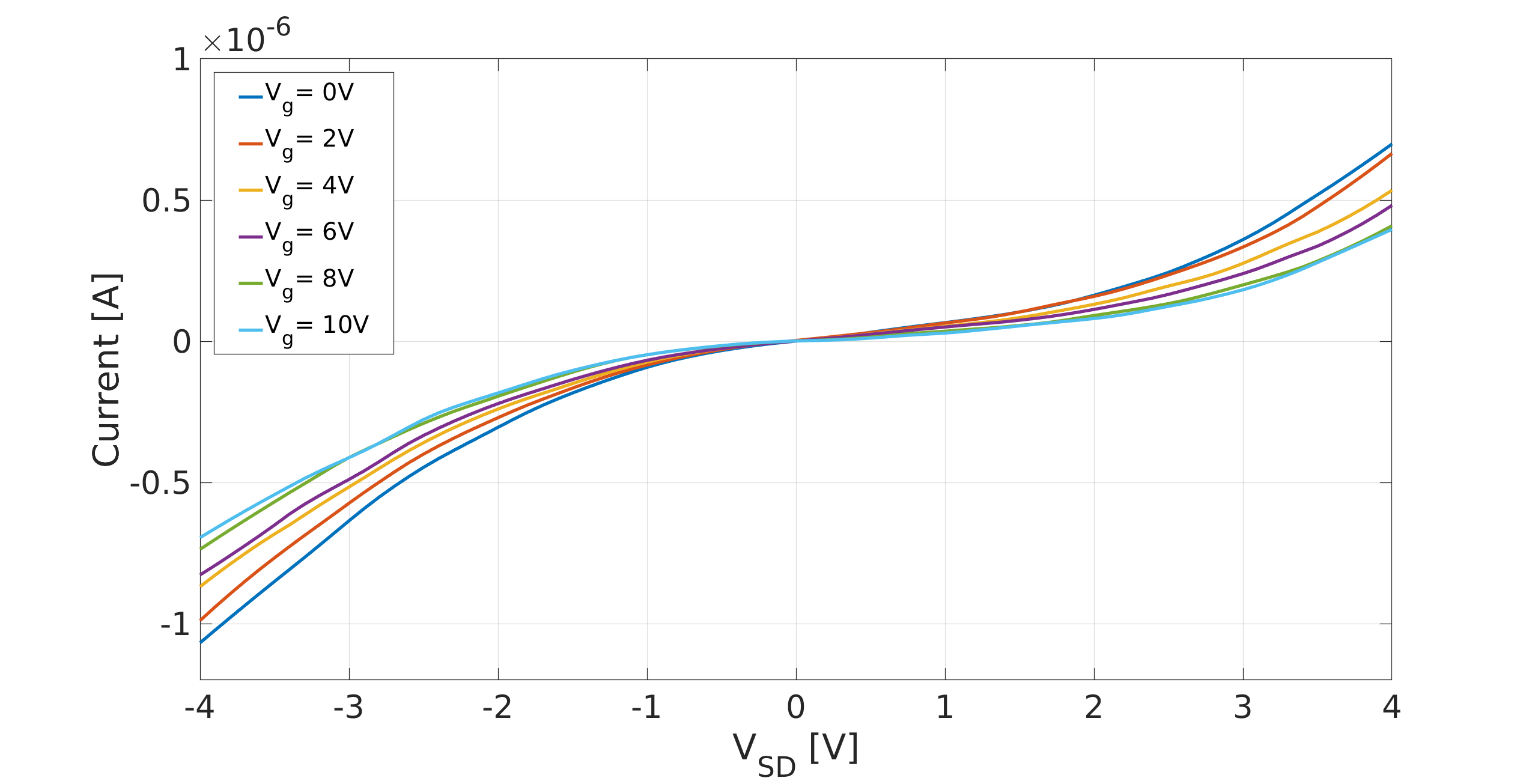}
 	\caption{I-V characteristics of fabricated device with 2 micron gap electrodes with $\unit{50}{KHz}$ frequency. }
 	\label{fig12}
 \end{figure}

\begin{figure}[t!]
	\centering
	\includegraphics[width=10cm,height=5.5cm]{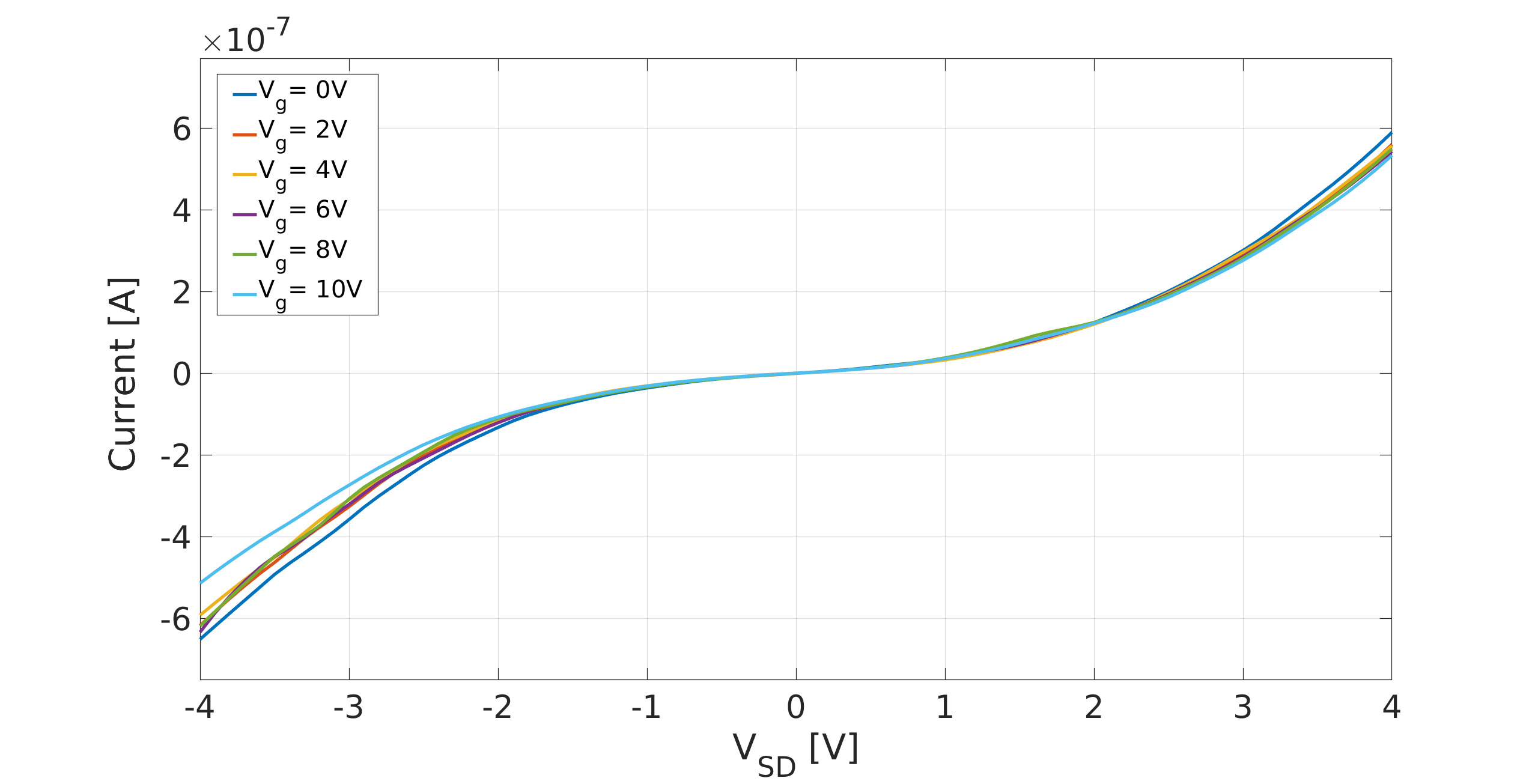}
	\caption{I-V characteristics of a fabricated device with 2-micron gap electrodes with $\unit{50}{KHz}$ frequency. }
	\label{fig14}
\end{figure}

The duration of the DEP process in all experiments was 20 seconds. Figure \ref{fig10} and Figure \ref{fig11} show the I-V curve of SWCNTs, respectively with 1 and 2-micron gap electrodes as a function of different source-to-drain voltages. Here, the
curves are related to the different gate voltage. High linearity of curves shows that most deposited CNT(s) are
metallic (low bandgap semiconductor) \cite{xue2011dielectrophoretic}. Figure \ref{fig12} and Figure \ref{fig14} show the current for several $V_{\text{g}}$ and $V_{\text{SD}}$ voltages in two different fabricated devices with 2 micron gap electrodes with $\unit{50}{KHz}$ frequency. Here, the deviation from linearity shows that aligned SWCNT(s) are
semiconductor \cite{collins1998nanoscale}.

 An important point to notice is that our model is just for a single SWCNT but experimental results are I-V characteristic of SWCNT bundles so we see a back gate response in I-V curves and very little deviation beginning at high source-drain
voltage. In this frequency interval, we align bundles of metallic semiconductor SWCNTs with dominant metallic
(low bandgap semiconductor) behavior. In fact, high linearity of curves shows that most deposited CNT(s) are metallic.

Here, we take the I-V characteristics of the SWCNT-FET into account in terms of linearity. In the case of semiconductor,
intrinsic, p-type or n-type, free of external voltage, thermal agitations of electrons in the valance band causes a continues
creation of electron-hole pairs i.e. electrons move from valence band to conduction band with a lifetime and result is a
a constant concentration of free electrons and holes at a constant temperature. Now using SWCNT as a channel in a
CNT-FET, the shape of the current-voltage characteristic is strongly affected by the potential profile across the
channel \cite{datta2005quantum}. Using a self-consistent field method  with increasing source-drain voltage we
get more states available for conduction and higher electron (hole) density in the conduction band in the case of a
semiconductor so increasing external voltage we get an increasing concentration of electrons and holes (up to a point
which is not present in our curves) and consequently continues increase in conductivity which results in the deviation
from linearity in the case of semiconductor SWCNTs \cite{datta2005quantum}.
 
 \begin{figure}[t!]
 	\centering
 	\includegraphics[width=10cm,height=5.5cm]{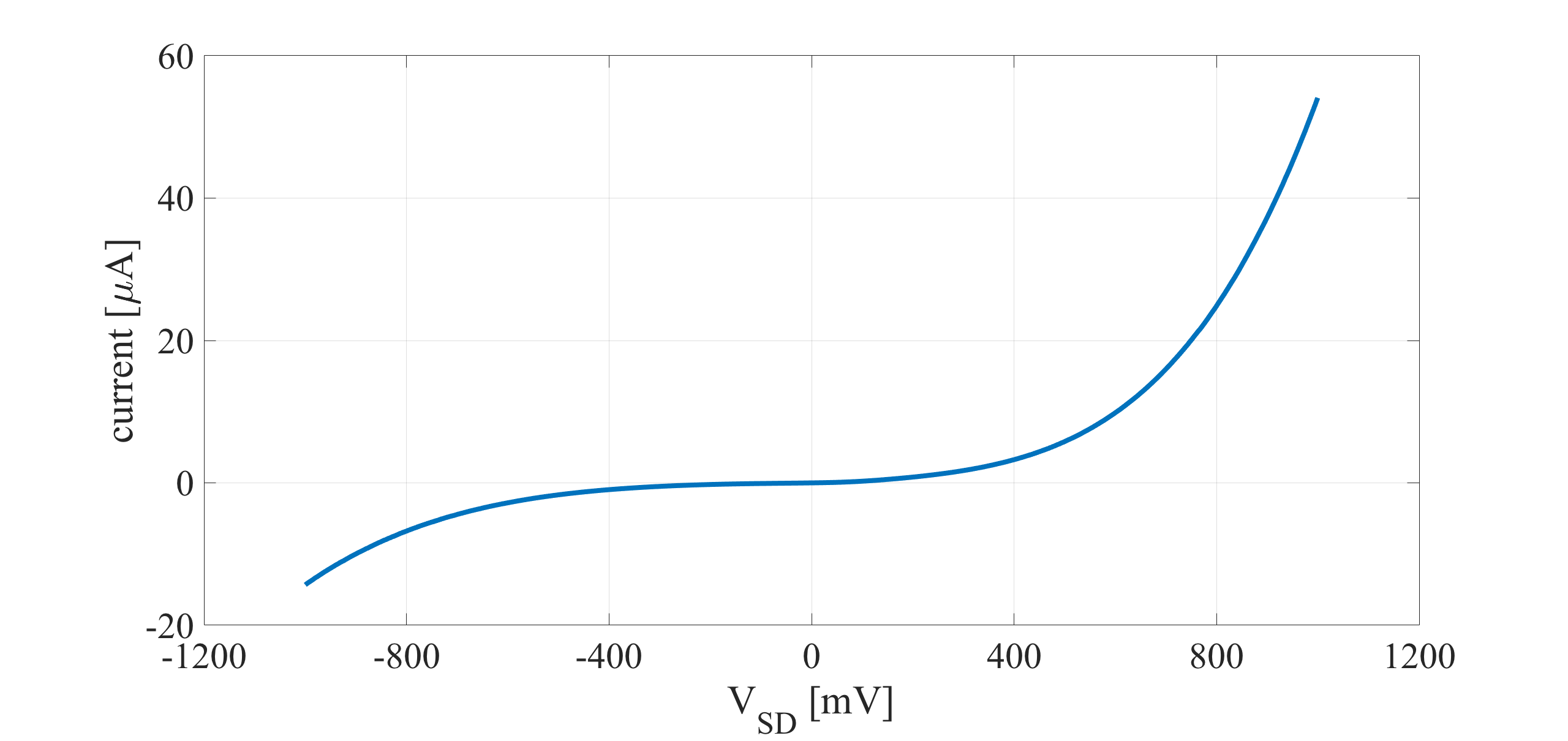}
 	\caption{I-V characteristics of fabricated device with 1 micron gap electrodes with 50KH frequency and $V_{\text{g}}=\unit{0}{V}$.
 		The asymmetry of the I-V curve originates from two different ohmic contacts.}
 	\label{fig15}
 \end{figure}

Asymmetry apparent in the curves originates from different Ohmic contacts of SWCNTs with metallic electrodes at two
endpoints. As Figure \ref{fig15} shows with a symmetric voltage applied the resulting I-V curve is not a symmetric one which as
stated is because of different contacts at the two endpoints.
\subsection{The SEM images}
Here we bring scanning electron microscope images of some fabricated devices. On the right side pictures, some ropes of carbon nanotubes consisting of one or more single-walled carbon nanotubes are apparent. Another point worth mentioning is the different currents at the same voltage in I-V characteristic curves of fabricated devices which seems to be because of the different number of SWCNTs aligned between electrodes in different devices. 
 
 \begin{figure}[ht!]
 	\centering
 	\subfloat{\includegraphics[width=0.45\textwidth]{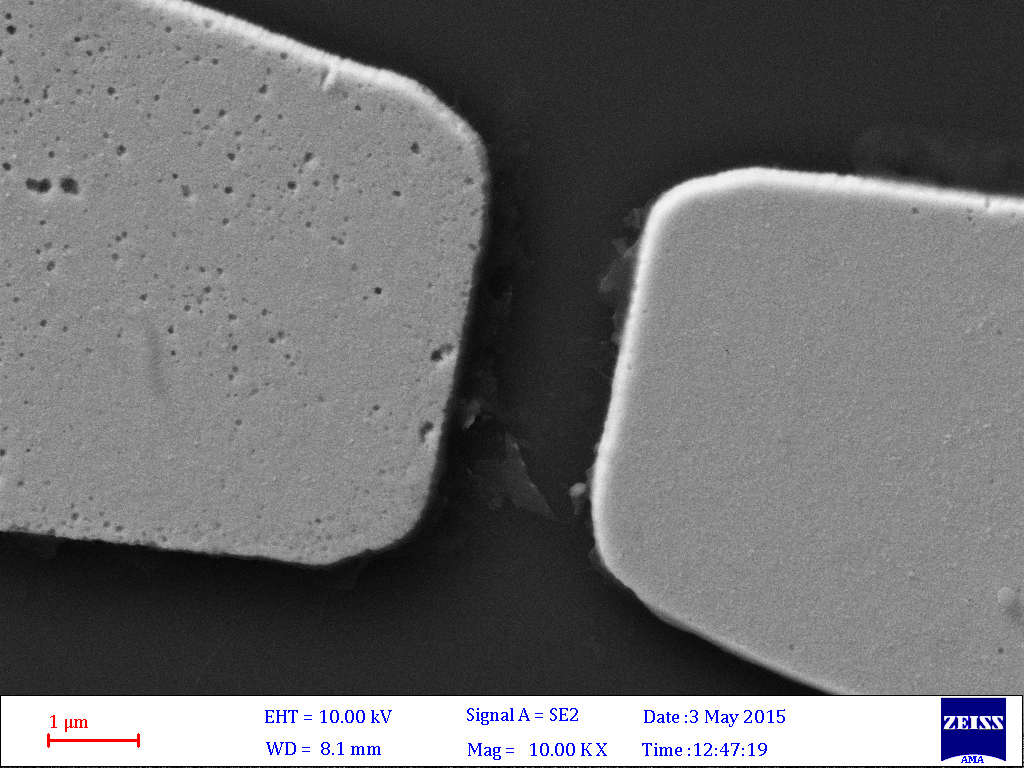}}
 	\hfill
 	\subfloat{\includegraphics[width=0.45\textwidth]{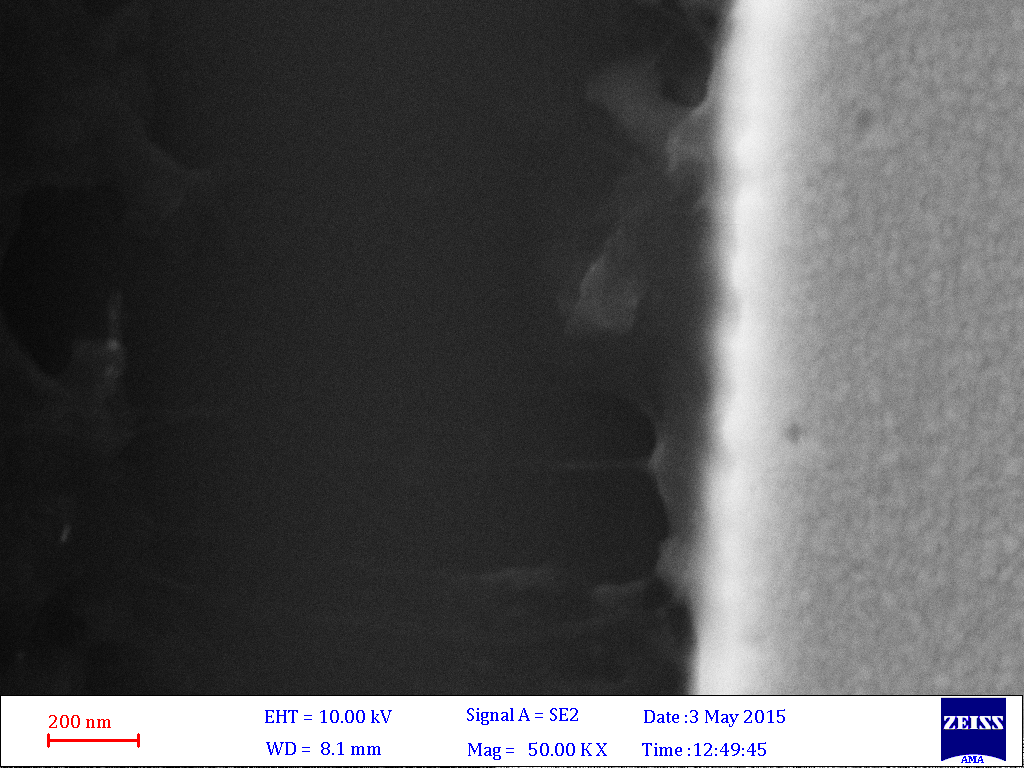}}
 	\caption{Left: SEM Images of the fabricated device related to the I-V curve of Figure \ref{fig12}. Right: the picture shows some ropes of carbon nanotube between two electrodes.}
 	\label{fig17}
 \end{figure}

 \begin{figure}[ht!]
	\centering
	\subfloat{\includegraphics[width=0.45\textwidth]{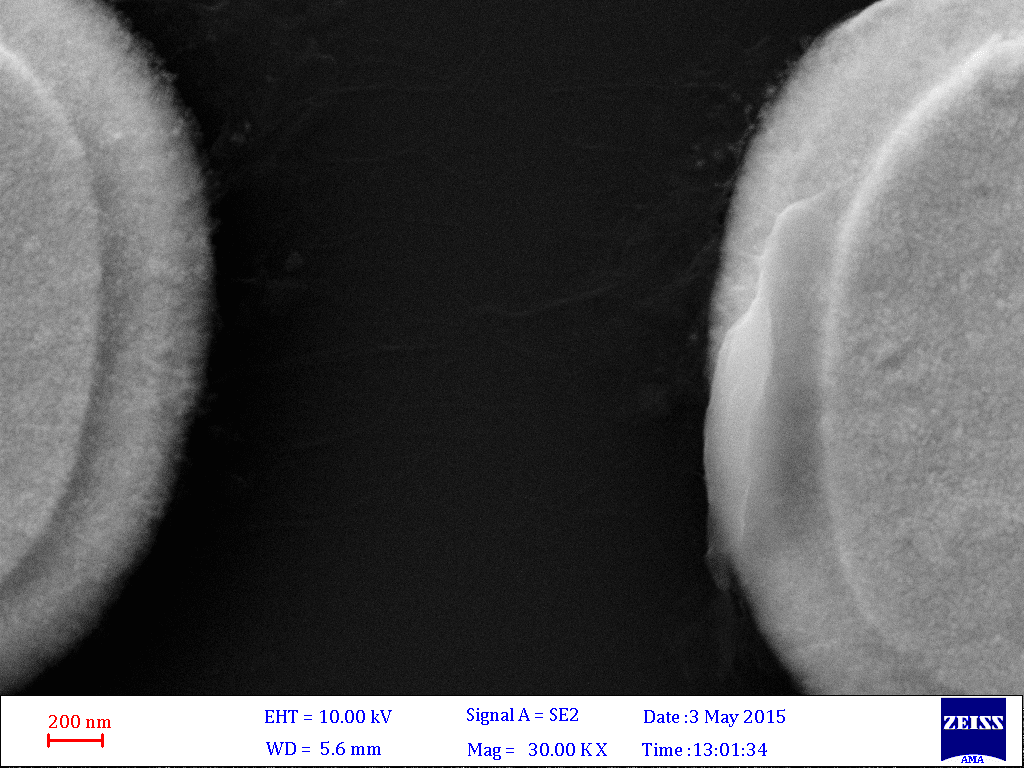}}
	\hfill
	\subfloat{\includegraphics[width=0.45\textwidth]{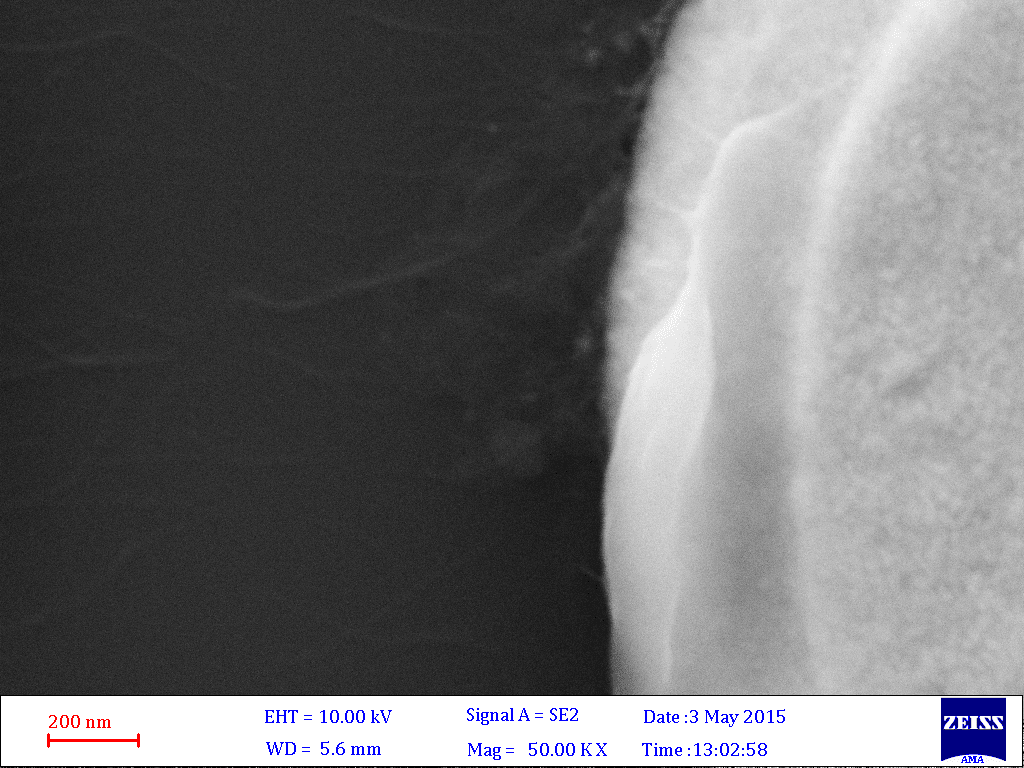}}
	\caption{Left: SEM Images of the fabricated device related to the I-V curve of Figure \ref{fig12}. Right: the picture shows some ropes of carbon nanotube between two electrodes.}
	\label{fig18}
\end{figure}

 \begin{figure}[ht!]
 	\centering
 	\subfloat{\includegraphics[width=0.45\textwidth]{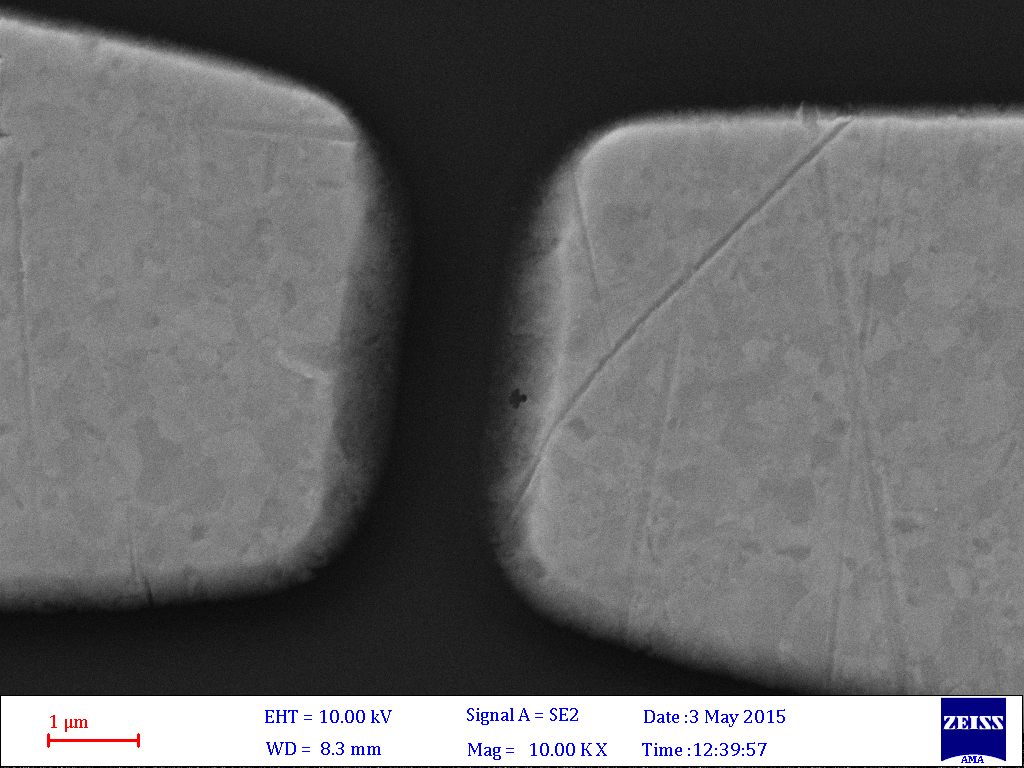}}
 	\hfill
 	\subfloat{\includegraphics[width=0.45\textwidth]{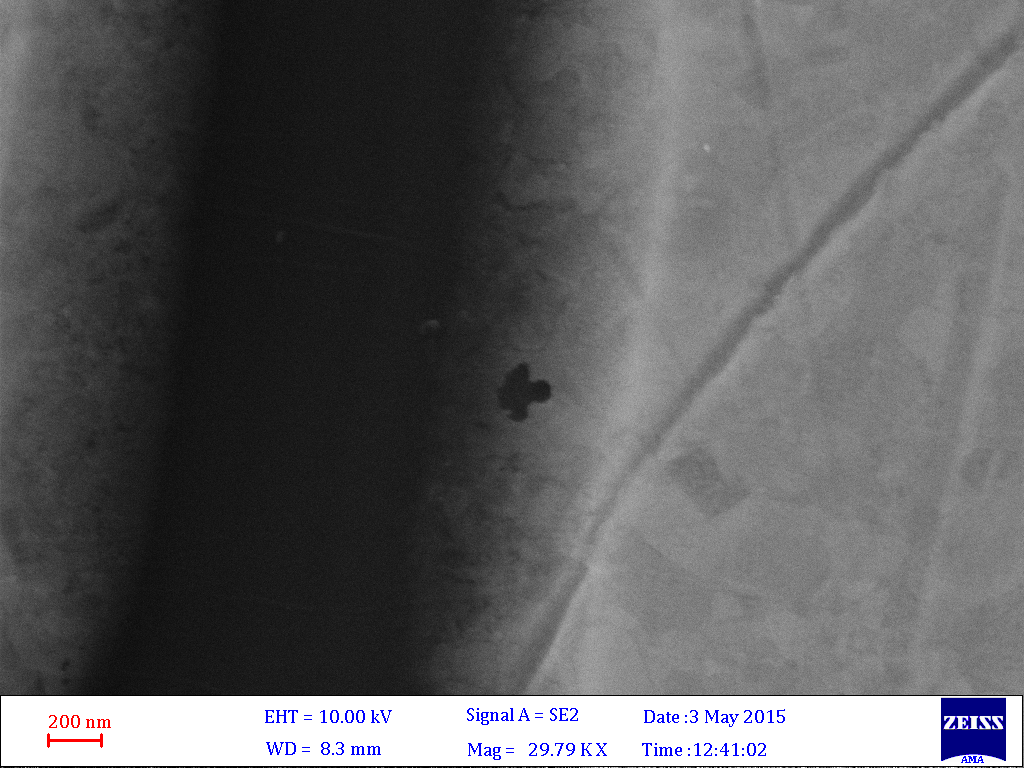}}
 	\caption{Left: SEM Images of the fabricated device related to the I-V curve of Figure \ref{fig15}. Right: the picture shows some ropes of carbon nanotube between two electrodes.}
 	\label{fig18}
 \end{figure}
 
\section{Conclusions}
\label{conclusions}
Experimental results show that the right frequency interval in the DEP process for alignment of single-walled carbon
nanotubes is the one suggested by the ellipsoid model and in the case of large gap semiconductor SWCNTs shows a great
difference; two orders of magnitude; with a spherical model which is prevalent in the articles. Picking the right frequency
interval, as the theoretical model suggests, adjusting other parameters of the process such as the geometry of electrodes
and changing the metal electrodes we can fully realize a more effective carbon nanotube field-effect transistor based on
the DEP process. It is also well depicted that, choosing the right parameters, dielectrophoresis can be applied as a reliable
way for carbon nanotube FET fabrication on a large scale.

\section{Acknowledgments}     
A. Khodadadian and C. Heitzinger acknowledge financial support given by the FWF (Austrian Science Fund) START project No. \textit{Y660 PDE Models for Nanotechnology}.

%

 
 \bibliographystyle{elsarticle-num}
 \bibliography{MLMC}
\end{document}